\pdfoutput=1 
\documentclass{JINST}
\usepackage{tabularx}
\usepackage{acronym}
\usepackage{subfigure}
\usepackage{graphicx}
\usepackage[english]{babel}
\usepackage{subfloat}
\usepackage{cite}
\usepackage{caption}
\usepackage{lineno}
 \usepackage{amsmath}
\usepackage{subfloat}

\makeatletter
\let\@fnsymbol\@arabic
\makeatother

\let\ifpdf\relax

\title{Results from the October 2014 CERN test beam of LumiCal\footnote{Talk presented at the International Workshop on Future Linear Colliders (LCWS15), Whistler, Canada, 2-6 November 2015.}\\
}

\author{ 

O.~Borysov$^a$, V.~Ghenescu$^b$, A.~Levy$^a$, I.~Levy$^a$,  S.~Lukic$^c$, J.~Moron$^d$,  A.T.~Neagu$^b$,  T.~Preda$^b$, O.~Rosenblat$^a$\\
\centerline {\bf{\Large{\rm{On behalf of the FCAL collaboration}}}}\\ \\
\llap{$^a$}Tel Aviv University, Tel Aviv, Israel\\
\llap{$^b$}ISS, Bucharest, Romania\\
\llap{$^c$}Vinca Institute of Nuclear Sciences, University of Belgrade, Serbia\\
\llap{$^d$}AGH University of Science and Technology, Cracow, Poland\\

E-mail: \email{levyaron@post.tau.ac.il}
 }

\abstract{
A prototype of a luminometer, designed for a future $e^+e^-$ collider detector, was tested in the CERN PS 
accelerator T9 testbeam.
The objective of this test beam was  
to demonstrate a multi-plane operation,
to study the development of the electromagnetic shower 
and to compare it with MC simulations. 
 }

\begin{document}

\section{Introduction}
\label{intro}

Two special calorimeters~\cite{ilc1} are foreseen in the very forward region of a detector for a future $e^+e^-$ linear collider experiment,
the Luminosity Calorimeter (LumiCal) and the Beam Calorimeter (BeamCal). 
The LumiCal will measure the luminosity with a precision of better than 10$^{-3}$ at 500~GeV centre-of-mass 
energy and 3$\times 10^{-3}$ at 1~TeV centre-of-mass energy at the ILC, and with a precision of 10$^{-2}$ at CLIC up to 3~TeV. 
The BeamCal will perform 
a bunch-by-bunch estimate of the luminosity and, supplemented by 
a pair monitor, assist beam 
tuning when included in a fast feedback system~\cite{grah_sapronov}.

\begin{figure}[h!]
  \centering
   \includegraphics[width=0.55\textwidth]{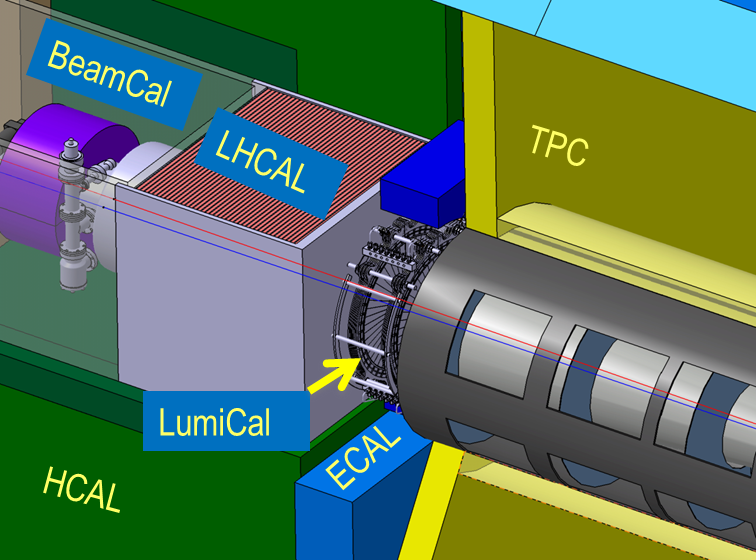}
\caption{The very forward region of the ILD detector. 
LumiCal, BeamCal and LHCAL are carried by 
the support tube for the final focusing quadrupole QD0 and the beam-pipe. 
TPC denotes the central tracking chamber, ECAL the electromagnetic and 
HCAL the hadron calorimeter. }
\label{fig:Forward_structure}
\end{figure}

Both calorimeters extend the detector coverage to low polar angles, 
important e.g. for new particle searches with a missing energy signature~\cite{drugakov}. 
A sketch of the design is shown in Figure~\ref{fig:Forward_structure} for the ILD detector. 
The LumiCal is positioned in a circular hole of the end-cap electromagnetic calorimeter ECAL.
The BeamCal is placed just in front of the final focus quadrupole.
LumiCal covers polar angles between 31 and 77~mrad and BeamCal, between 5 and 40~mrad.

Both calorimeters consist of 3.5~mm-thick tungsten absorber disks, each corresponding
to around one radiation length, interspersed with sensor layers. Each sensor layer is segmented radially
and azimuthally into pads. The read-out rate is driven by the beam-induced background.
Due to the high occupancy originating from beamstrahlung and two-photon processes,  
both calorimeters need a fast readout. Front-end (FE) and ADC ASICs are placed at the outer radius of the
calorimeters. 
In addition, the lower polar-angle range of BeamCal is exposed to a large flux 
of low energy electrons, resulting in depositions up to one 
MGy for a total integrated luminosity of 500~fb$^{-1}$ at 500~GeV. Hence, radiation-hard sensors are needed.
Prototype detector planes assembled with FE- and ADC-ASICs for 
LumiCal and for BeamCal have been built. In this paper, results of the 
performance of a prototype of LumiCal, following tests in the CERN PS beam, are reported. 

\section{Overview of testbeam instrumentation}
The performance of fully instrumented LumiCal and BeamCal detector planes was studied in previous testbeam
campaigns. The results were fully matching the requirements~\cite{TB11}. The next step 
in the detector prototype development was to prepare and conduct a 
testbeam study of a multi-plane structure, 
performed in October 2014 at the T9 east area of the proton synchrotron (PS) at CERN.

\subsection{Mechanical framework}\label{ch2_sec:mechanical}

\begin{figure}[!ht]
		\centering
		\begin{minipage}{.525\columnwidth}{
				\includegraphics[width=\columnwidth]{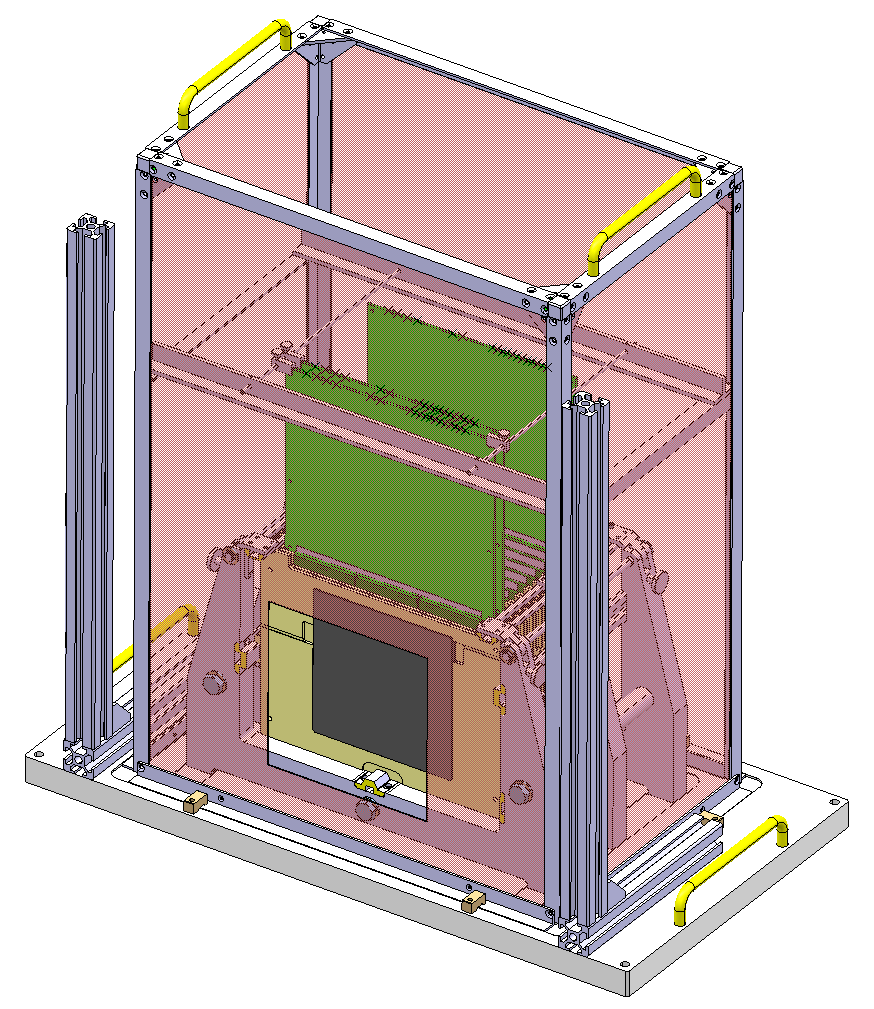}
			}
		\end{minipage} \qquad
		\begin{minipage}{.4\columnwidth}
			{
				\includegraphics[width=\columnwidth]{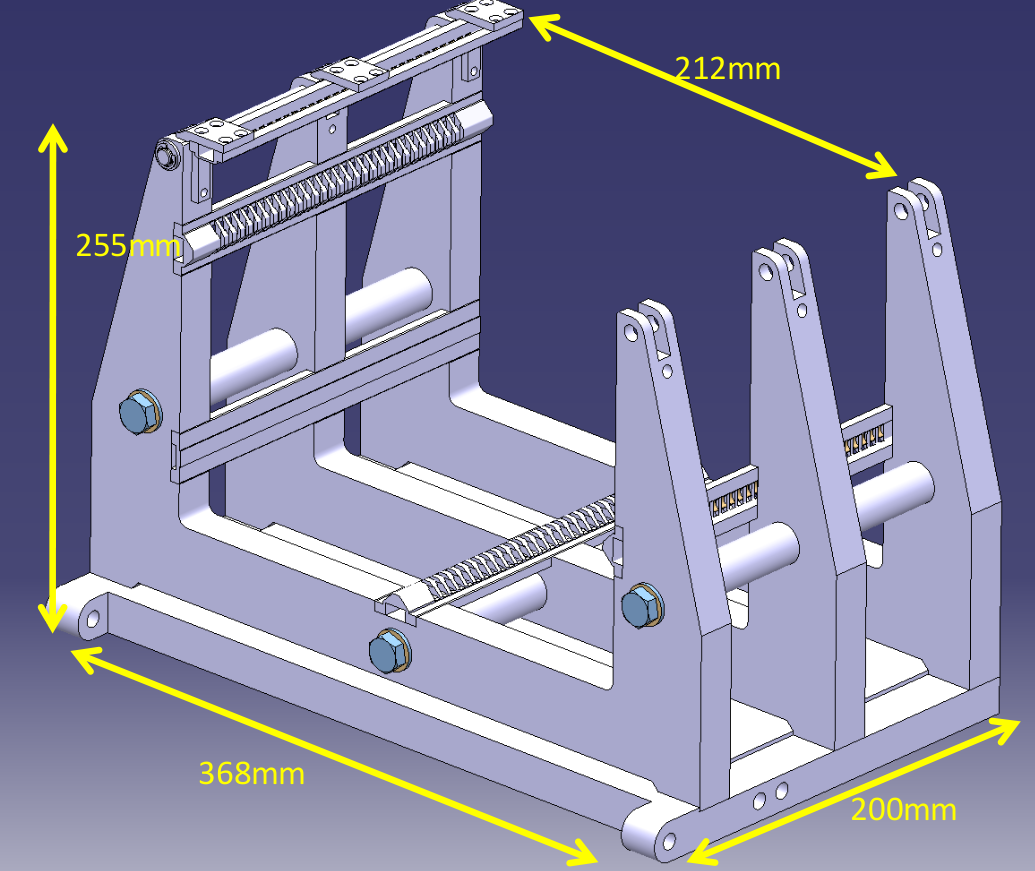}
			} \qquad \qquad
		{
				\includegraphics[width=\columnwidth]{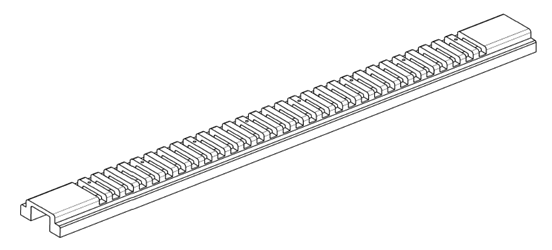}
			}
		\end{minipage}
		\caption{Left-hand side: Scheme of the complete mechanical structure with the tungsten 
  absorbers and readout boards installed. 
Upper right-hand side: Dimensions of the precision mechanical frame for the positioning of sensor and absorber planes. 
Lower right-hand side: Detail of the retaining comb jig for positioning of sensor and absorber planes.}
\label{ch2_fig:lumi_mech_overview}
	\end{figure}

To allow the multiple-plane operation, a sophisticated mechanical structure to meet the demanding geometrical requirements
was developed at CERN~\cite{nuiry}. 
The required precision of the shower polar angle reconstruction imposes a  
precision in the positioning of the sensors of a few tens of micrometers. 
Since only four detector planes were available, the mechanical structure had to 
enable modifications in the prototype layout during the testbeam, 
allowing to measure the longitudinal shower profile. 
The overall view of the structure of the mechanical frame is presented in 
Figure~\ref{ch2_fig:lumi_mech_overview}. The most important component, the layer positioning structure,  includes three aluminum combs with 30 slots each,  which allow to install the tungsten absorber or active sensor layer with the required precision. 

\subsection{Permaglass frames}

\begin{figure}[!ht]
		\centering
		{
			\includegraphics[width=0.45\columnwidth]{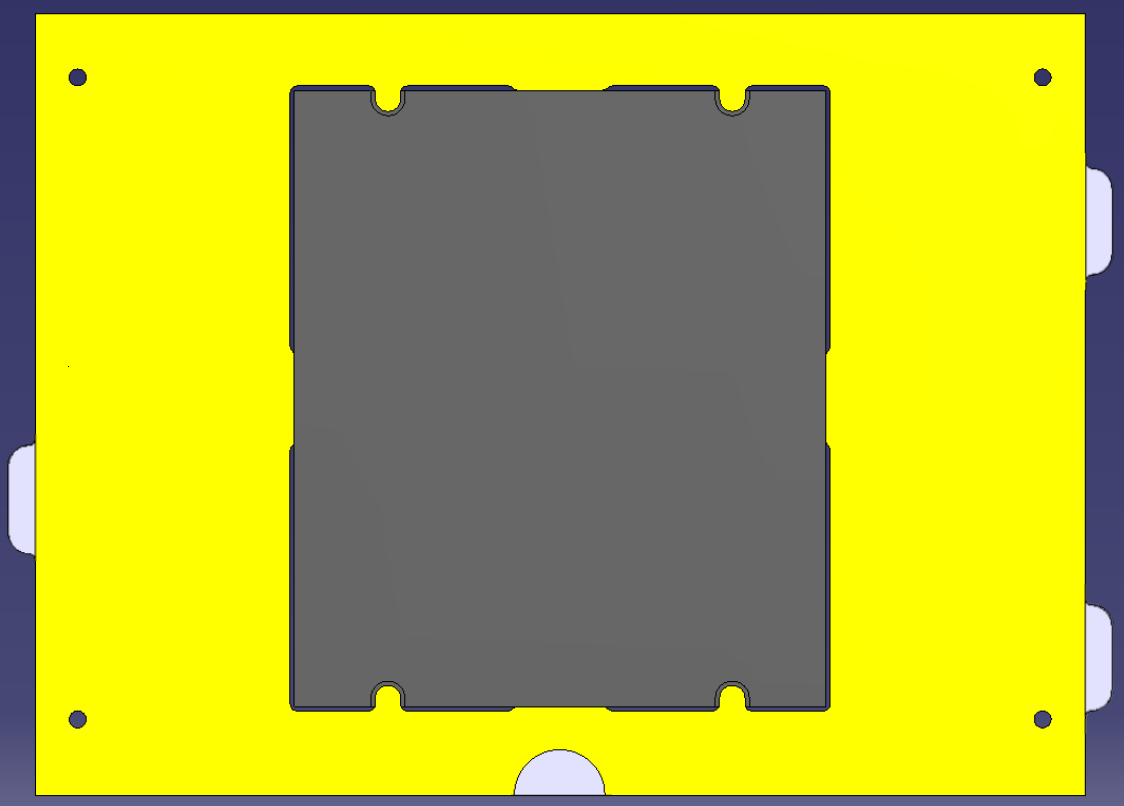}
		} \qquad
		{
			\includegraphics[width=0.45\columnwidth]{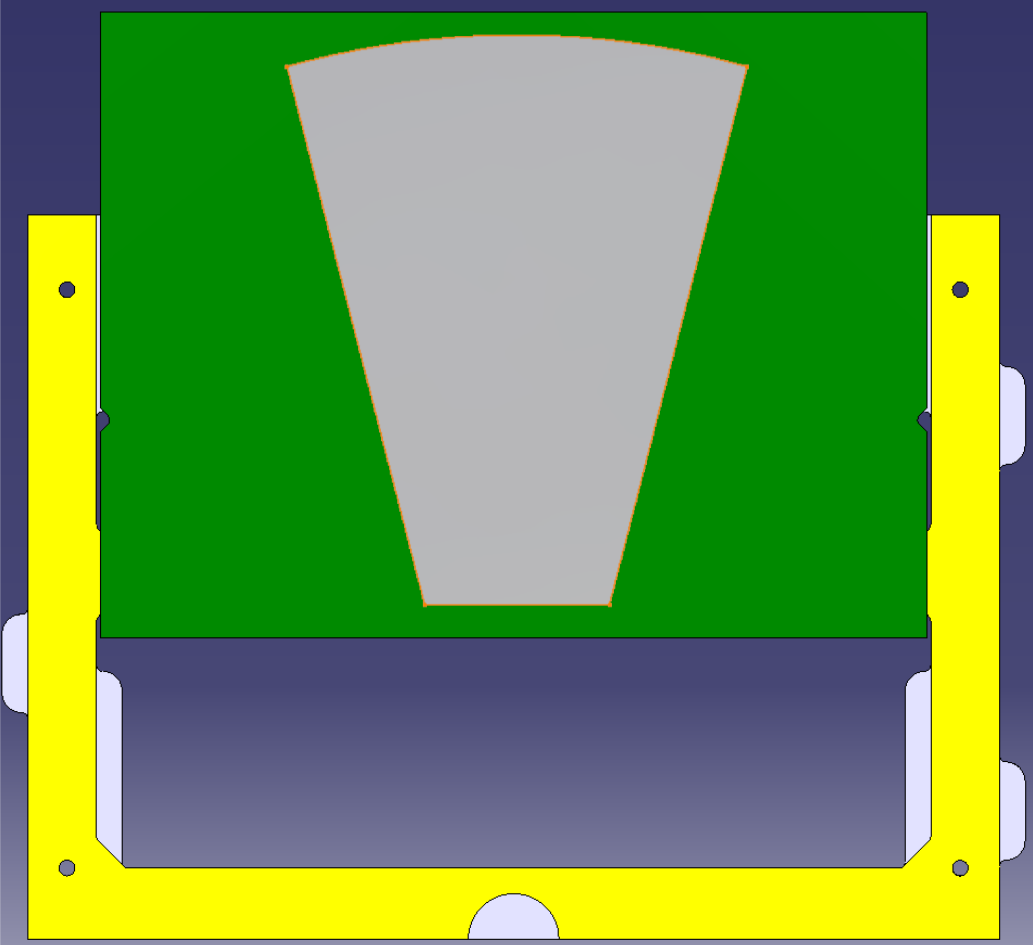}
		}
		\caption{Detector layers supporting frames. Left-hand side: tungsten absorber (gray) in a 
              permaglass frame (yellow). Right-hand side: sensor (gray) PCB (green) partially retracted from a permaglass 
             frame (yellow).}
		\label{ch2_fig:permaglass_frames}
	\end{figure}

\begin{figure}[!hb]
		\centering
		\includegraphics[angle=-90,width=.8\columnwidth]{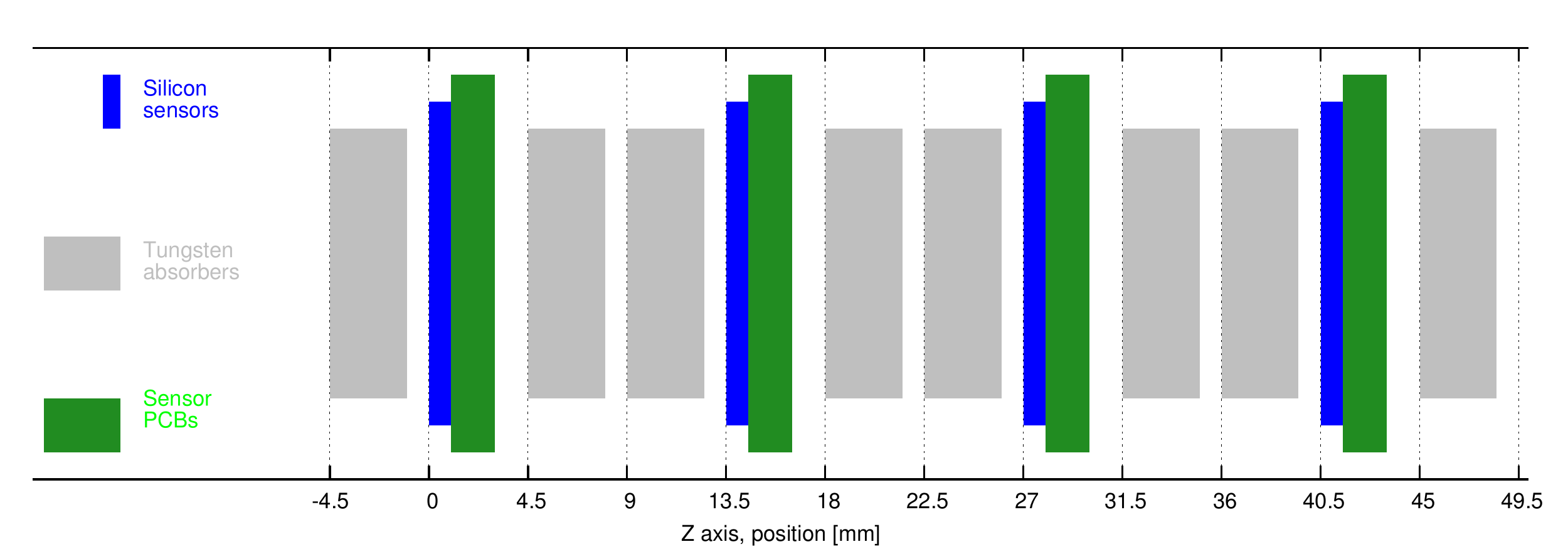}
		\caption{Geometry of the first configuration of the LumiCal detector prototype with active sensor layers located in place of tungsten absorbers. The Y axis (perpendicular to Z) is not to scale.}
		\label{ch2_fig:lumi_geom_real}
	\end{figure}

The 3.5~mm thick tungsten absorber plates are mounted in permaglass frames, as shown in Figure~\ref{ch2_fig:permaglass_frames}. 
The light gray inserts located on the left, right (outside) and bottom (inside) 
sides of the permaglass frames (yellow) contain small bearing balls 
providing precisely positioned support points for the comb slots.
Currently the sensors are mounted onto a 2.5~mm thick PCB\footnote{this thickness will be reduced in the future to be less than 1 mm using 
a sophisticated connectivity scheme under development. 
Then sensor planes will be positioned in the 1 mm gap between absorber planes.} serving as the mechanical 
support and high voltage supply. 
Therefore, sensor layers are positioned in slots foreseen for tungsten absorber planes to form a stack, as shown in 
Figure~\ref{ch2_fig:lumi_geom_real} for the first configuration. Different configurations of absorber and sensor planes can be made to measure electromagnetic showers 
at several positions inside the stack. 

The positioning precision of the supporting structure, as well as the tungsten absorber thickness uniformity, 
were extensively tested~\cite{nuiry}. The maximum differences between the theoretical and 
measured absorber plane positions do not exceed the required $\pm$50~$\mu$m precision\footnote{This 
requirement will become relevant when thin sensor planed will be used in the future.}. 
A function test of the fully assembled stack was performed before installation in the testbeam and the full detector prototype functionality was confirmed.

\subsection{Detector module geometry}\label{ch2_sec:geometry}

For 5~GeV electrons from the PS accelerator beam the expected maximum of the shower is located around the 6th absorber layer. 
In order to measure the full shower development, data should be taken with at least the first 10 layers instrumented. 
Since only four readout boards are available, this condition could be met only with at least three absorber layers 
placed between each active sensor layer. 
However, this leads to a rather small number of measurements which could result in large uncertainties in the shower shape. 
Since the mechanical support structure enables relatively simple detector geometry changes, a different approach was applied.
Three different detector configurations were used, with the active sensor layers always separated by two absorber layers. 
By adding additional absorber layers upstream the detector, the sensor layers were effectively moved downstream in the shower. 
The first stack configuration is shown in Figure~\ref{ch2_fig:lumi_geom_real}.
A summary of all configurations used is given in Table~\ref{ch2_tab:tb_conf}.   
The single absorber 
layer after the last silicon sensor was added in order to simulate the particles backscattering as it takes place 
in the target detector.
The shower can be therefore sampled up to the 10th layer with a shower sampling resolution of one radiation length.
Since the positions of sensors S1--S3 in the first configuration are replicated by the S0--S2 in the second one, 
the measurement results for the corresponding sensors from both configurations should be almost identical, being a lever arm to 
control response variation in the measurement. 
The data can be combined in order to imitate the detector prototype comprising nine active sensor layers.
	\begin{table}[h!]
		\centering
		\vspace{-0.3cm}
		\begin{tabular}{c|c|c|c|c|c|c|c|c|c|c}
			& \multicolumn{10}{c}{Radiation lengths in number of absorber layers} \\ \hline
			Configuration & 1 & 2 & 3 & 4 & 5 & 6 & 7 & 8 & 9 & 10 \\ \hline
			1 & S0 & \ \ \ \ \ & S1 & & S2 & & S3 & & & \\
			2 & & & S0 & & S1 & & S2 & & S3 & \\
			3 & & & & S0 & & S1 & & S2 & & S3 \\
		\end{tabular}
		\caption{Positions of active sensor layers in three configurations expressed in number of absorber layers (i.e. radiation lengths in absorber $X_0$) in front of the sensor layer. S0--S3 stands for Sensor 0 -- Sensor 3.}
		\label{ch2_tab:tb_conf}
	\end{table}
Due to a malfunctioning of the FPGA on the readout board of S3 in the third configuration, only eight positions are used in the analysis.

\section{Testbeam area instrumentation}\label{ch2_sec:testbeam_instrument}

\begin{figure}[!ht]
		\centering
		\includegraphics[width=0.95\columnwidth]{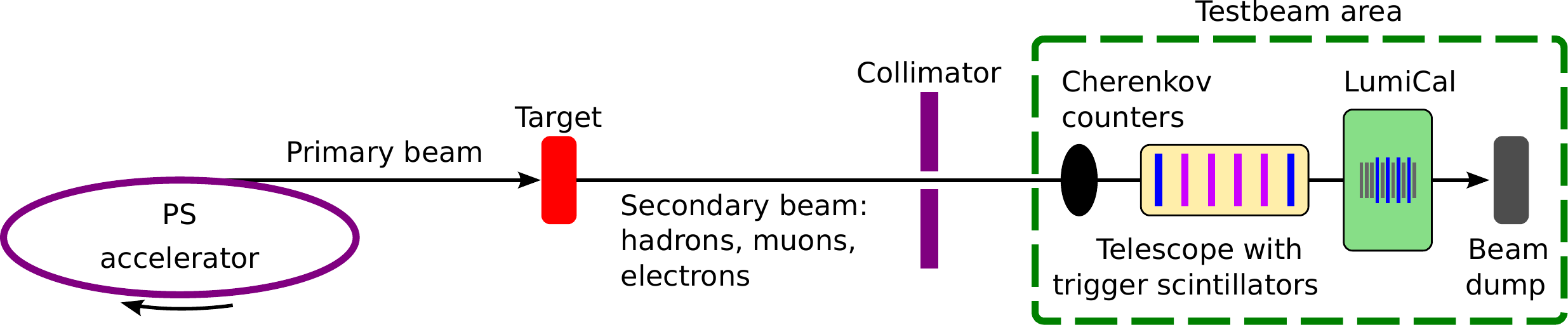}
		\caption{Overall, simplified view (not to scale) of the testbeam instrumentation.}
		\label{ch2_fig:tb_beam_path}
	\end{figure}

The PS accelerator provides a primary proton beam with momentum of 24~GeV/c. 
Since the beam is shared between different testbeam instrumentations and the LHC beam delivery system, 
the primary beam for the T9 area is provided in 400~ms long spills with a typical time separation of 33.6~s 
between them. The primary beam is converted using several targets and for the present testbeam it 
provided a secondary beam with muons, pions, hadrons and electrons with momenta in the range of 1--15~GeV/c. 
A narrow band of particle momenta centered at 5~GeV was selected using a dipole magnetic field and a set of collimators.
The simplified overall view of the PS east area testbeam 
facility is presented in~Figure~\ref{ch2_fig:tb_beam_path}.

\begin{figure}[!h]
		\centering
		\includegraphics[width=0.95\columnwidth]{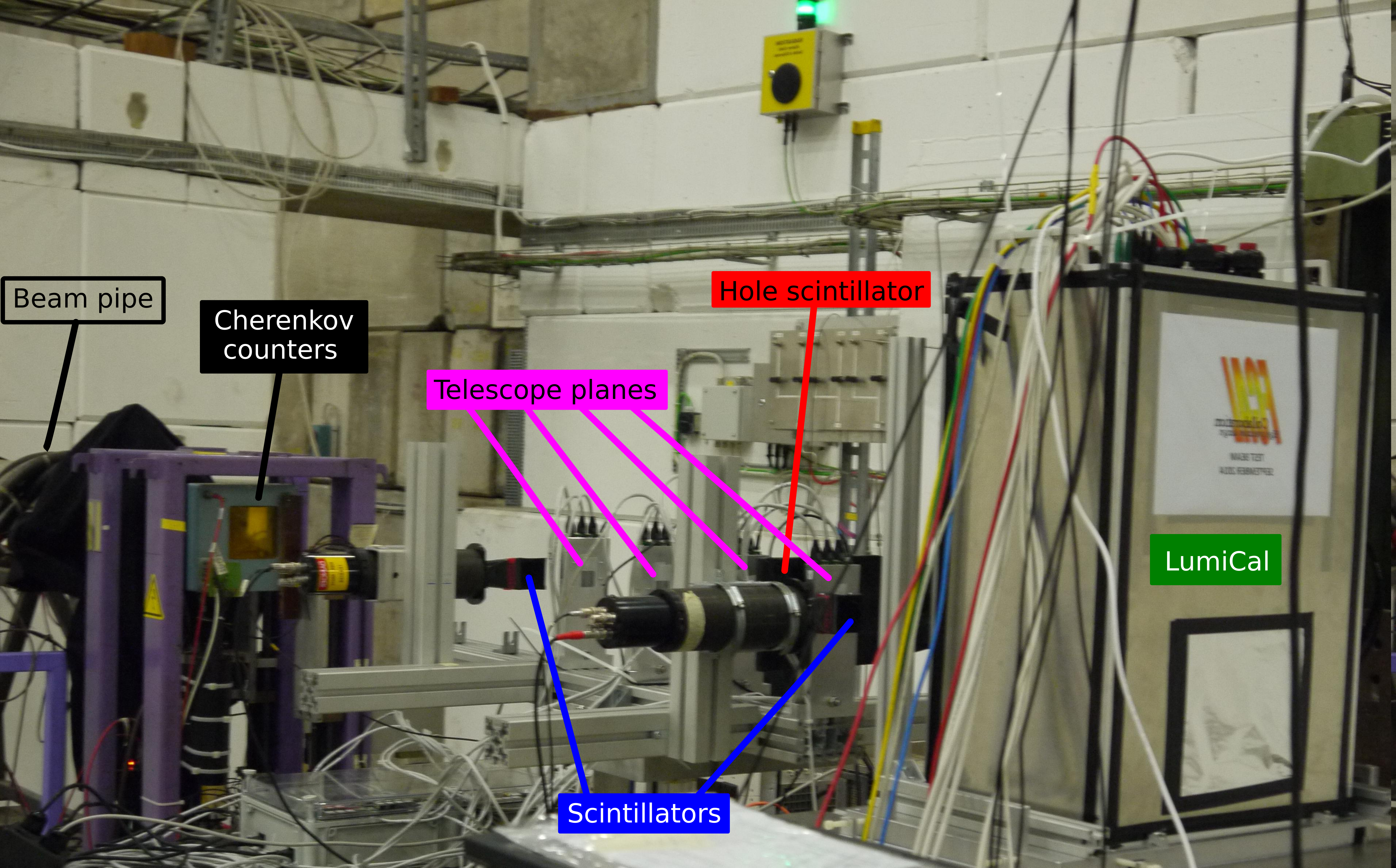}
		\caption{Overall photograph of the testbeam area instrumentation.}
		\label{ch2_fig:tb_area_pic_over}
	\end{figure}

\begin{figure}[!h]
		\centering
		\includegraphics[angle=-90,width=.8\columnwidth]{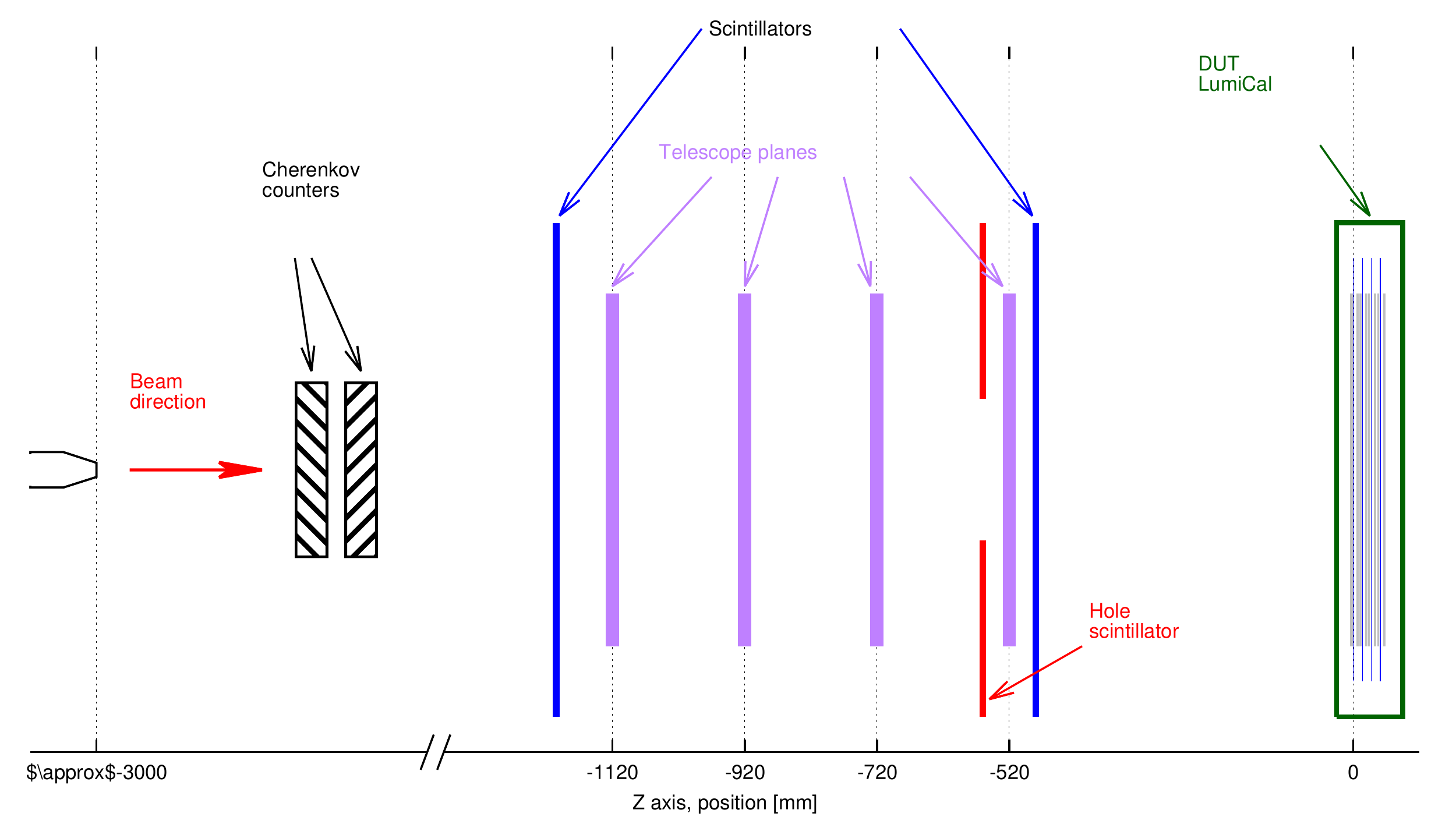}
		\caption{Testbeam area instrumentation geometry. Not in scale.}
		\label{ch2_fig:tb_area_geom}
	\end{figure}

\begin{figure}[!ht]
		\centering
		\includegraphics[width=0.7\columnwidth]{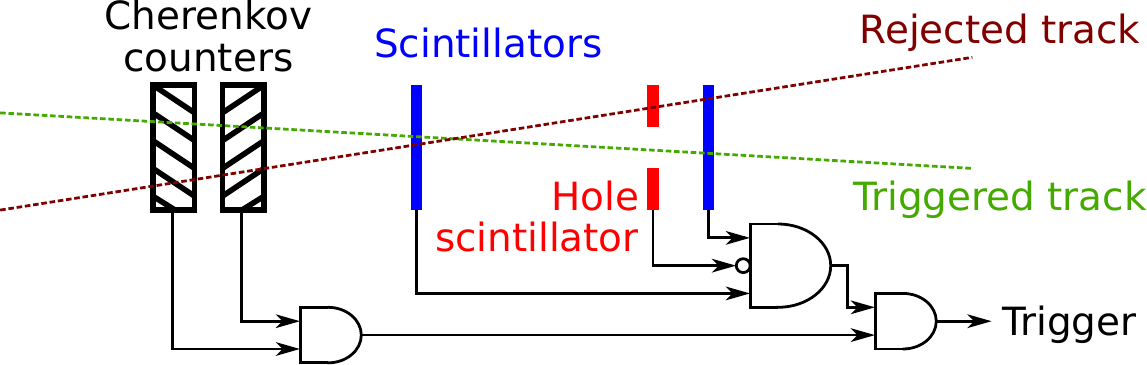}
		\caption{Scheme of the trigger system (see text).}
		\label{ch2_fig:tb_trigger_system}
	\end{figure}

A photograph showing the details of the testbeam area 
instrumentation is presented in~Figure~\ref{ch2_fig:tb_area_pic_over}. 
The schematic diagram of the instrumentation geometry is shown in~Figure~\ref{ch2_fig:tb_area_geom}. 
Since the secondary beam consists of a mix of various particles, the Cherenkov counters were used to 
discriminate the electrons or/and muons. The gas Cherenkov counters allow particle identification 
as well as the energy discrimination by changing the HV of the photomultiplier and the pressure of the CO$_2$. 
For an electron energy of 5~GeV, the gas pressure was set to 53~kPa with the HV set to 1900~V and 1950~V \
for the first and second Cherenkov counter, respectively. 
To reconstruct the trajectories of beam particles a multilayer tracking detector, the so-called telescope, 
developed by the Aarhus University, was used. The telescope utilizes the MIMOSA-26 chips, 
a monolithic active pixel sensor with fast binary readout~\cite{mimosa26}. 
One MIMOSA-26 chip comprises 1152~$\times$~576 pixels with 18.4~$\mu$m pitch, resulting in an active area of 21.2~$\times$~10.6~mm$^2$. The binary readout accepts the pixel signals exceeding a preset discrimination level. 
The pixel matrix is read continuously providing a complete frame every 115.2~$\mu$s. 
The data are gathered, triggered and stored by a custom DAQ system, based on the National Instrument PXI crate, 
developed by the Aarhus University in collaboration with the Strasbourg University. 
Four telescope planes, each comprising one MIMOSA-26 chip, were set upstream of the stack 
as shown in~Figure~\ref{ch2_fig:tb_area_geom}.
Three scintillation counters were used to provide a trigger for particles traversing the active part of the 
telescope sensors and the region of the sensors in the stack being read out. 
Two 5~$\times$~5~cm$^2$ scintillators tiles were placed upstream and downstream of the telescope 
(marked in blue in~Figure~\ref{ch2_fig:tb_area_geom}) and one (marked in red), with a 9~mm diameter circular hole, 
was placed just before the last telescope plane. 
Compact photomultipliers were attached to the scintillators providing electrical pulses.
In order to ensure that triggers are only generated by beam particles in the sensitive area
of the telescope, the signal from the hole scintillator was set in anti-coincidence.
The trigger signal was combined with the Cherenkov counters response, as shown in~Figure~\ref{ch2_fig:tb_trigger_system}, 
to create a trigger for electrons. 
A fraction of particles passing through the active area of the anti-coincidence scintilator are registered due to partial inefficiency of the scintilator.
An exemplary track of such particle is shown in Figure~\ref{ch2_fig:tb_trigger_system} as a dark-red line, 
labeled as a rejected track. 

The main limitation in the data statistics is the split into  400~ms long spills, containing typically 10$^3$--10$^4$ particles. 
Taking into account the spill repetition frequency, slightly below 2 per minute ($\approx$0.03~Hz), 
and the electron fraction of about 5~\%, a rate of 1.5--15 electrons per second was expected. 
In addition, the DAQ applies a veto mechanism rejecting triggers during the event data packaging. As a result, 
some of the electron triggers generated during the 400-ms long spill will be rejected, \
reducing the average event rate.  
Since the MIMOSA-26 chip provides a continuous readout, a high particle intensity 
during short spills can produce multiple track events, to be considered in the data analysis.

\section{Setup and data taking}

\begin{figure}[h]
\centering
\includegraphics[width=0.7\textwidth]{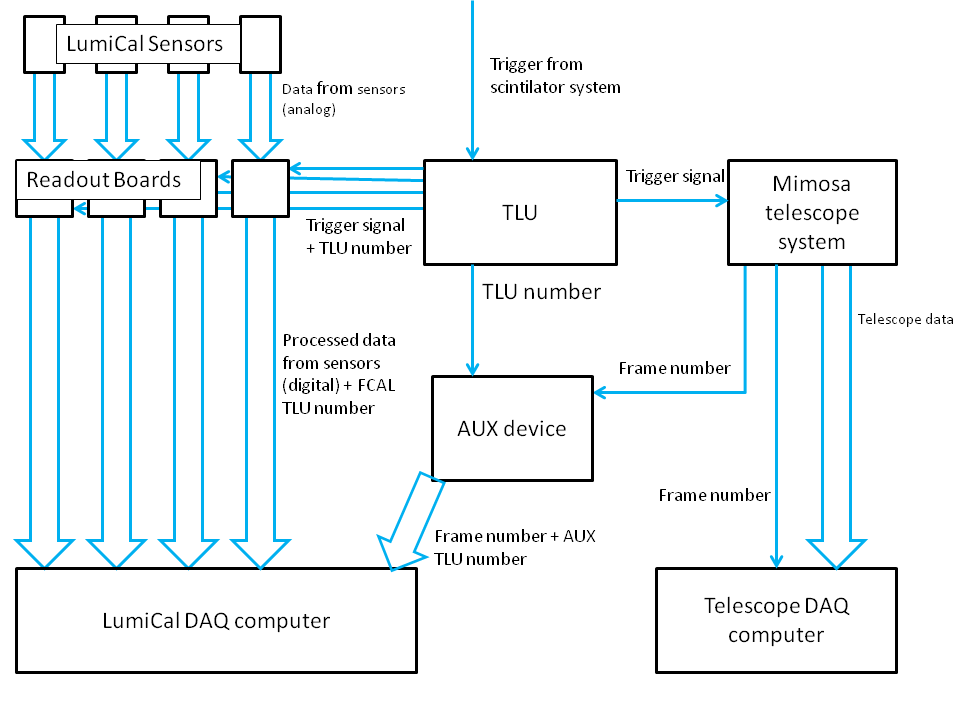}
\caption{A schematic of the connections and the path of the signal in the system. Data is symbolized by a thick arrow, while simple bit information (e.g. the trigger signal) is symbolized by a thin arrow.}
\label{fig:connections}
\end{figure}

There were two independent DAQ computers for the telescope and for LumiCal. 
In order to monitor and distribute the trigger signal correctly, a trigger logic unit (TLU)~\cite{TLU} was used. 
The TLU receives the trigger signal and generates an integer TLU number, 
counting the number of triggers received. The TLU then passes on the trigger signal and the TLU number to the telescope and the stack,
respectively. 
In order to coordinate between the TLU number and the telescope frame number a dedicated auxiliary (AUX) unit was used, 
saving these two numbers for the same event. Figure~\ref{fig:connections} displays the connections and 
the flow of information in the system.

\begin{figure}[!h]
 \centering
\includegraphics[width=0.45\textwidth]{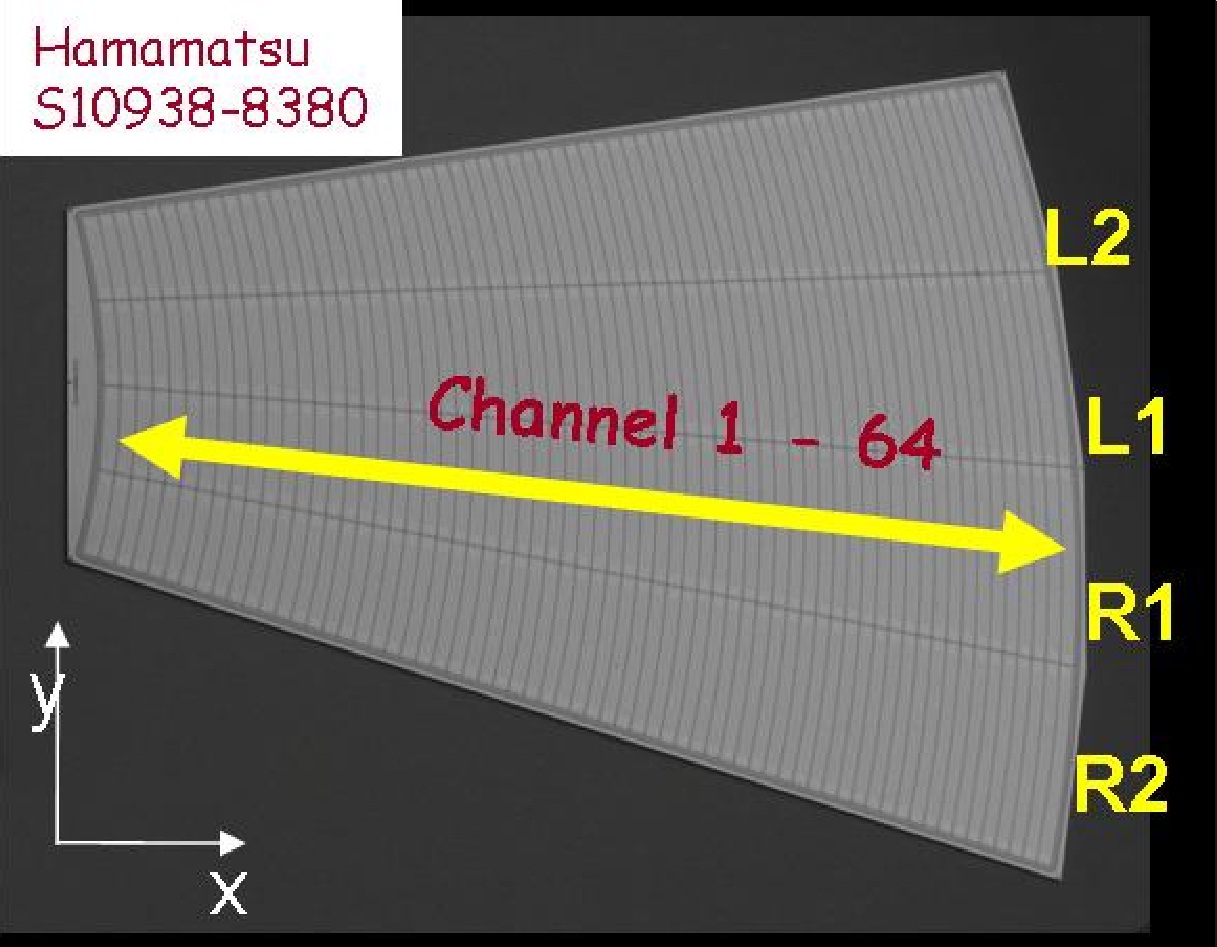}
\caption{A prototype silicon sensor for LumiCal.}
\label{figure:Lumical sensor}
\end{figure}

In each of the sensor planes, the connected pads were pads 51-64 of sector L1 and 47-64 of sector R1, as shown
 in Figure~\ref{figure:Lumical sensor}. 
Upon the arrival of a trigger,  the 32 channels were recorded with 32 samplings in the ADC in steps of  50 ns, 
resulting in a total read time of 1.6 $\mu s$. For each trigger, the TLU number from the LumiCal DAQ, the TLU number 
from the AUX and the telescope frame number from the AUX were stored 
for the purpose of synchronization between the stack and the telescope data.

\subsection{Telescope alignment and tracking}

Since for the given trigger and beam structure a single particle track per event in average was estimated, 
the expected particle hit count per event per telescope plane is around one.
However, the noise hits contribution is quite high and dominates over hits generated by the beam particles, 
especially for the fourth telescope plane. 

As the positions of the telescope planes were only roughly set by the telescope mechanical support structure, an 
alignment was required to define and correct for the testbeam telescope planes mutual positions.
This was done using the standard procedure implemented in TAF~\cite{TAF}.

\begin{figure}[!ht]
{
			\includegraphics[angle=-90,width=.49\columnwidth]{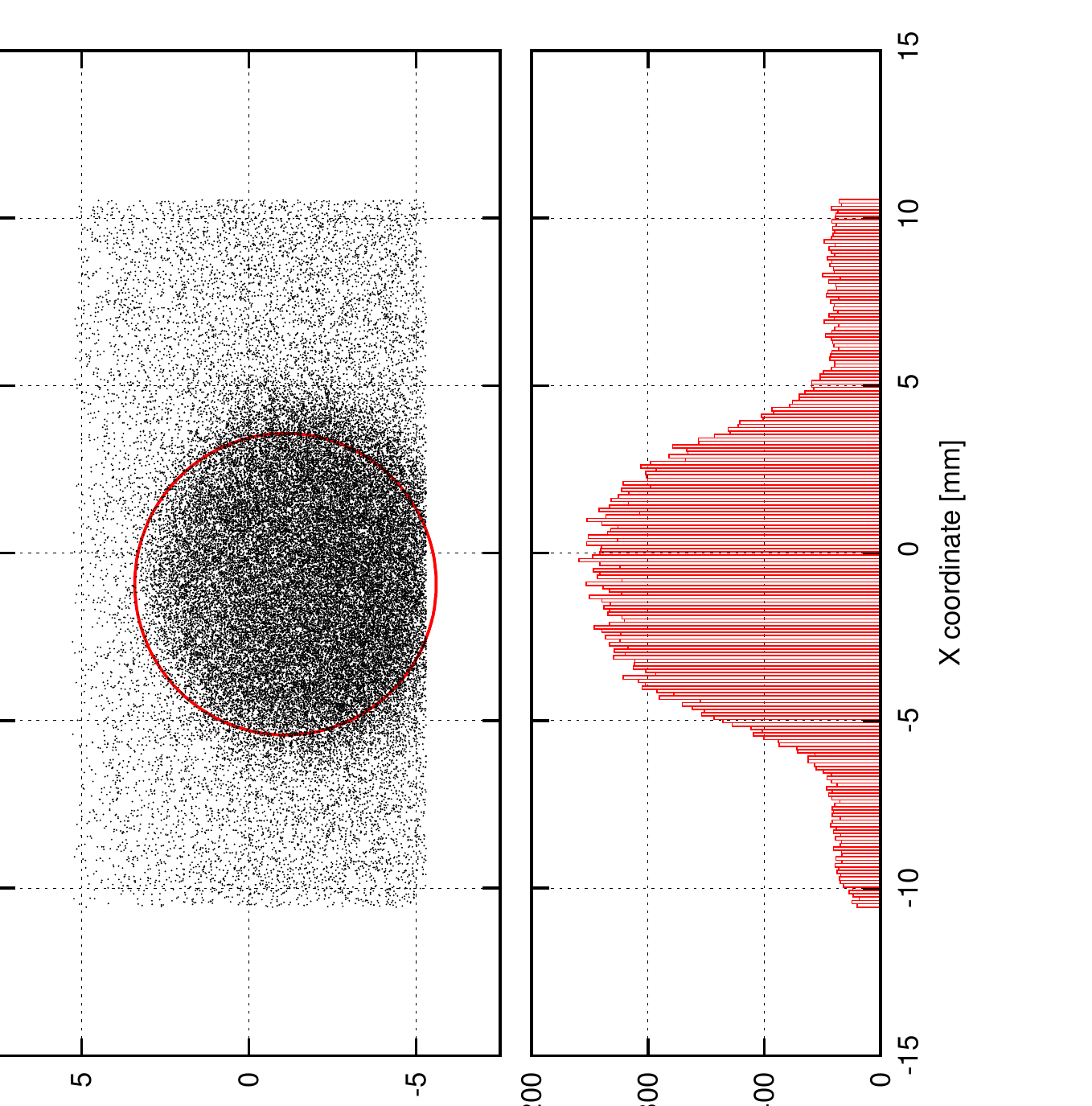}
			\label{ch2_fig:tracks_telemap_0}
		} \qquad
{
			\includegraphics[angle=-90,width=.49\columnwidth]{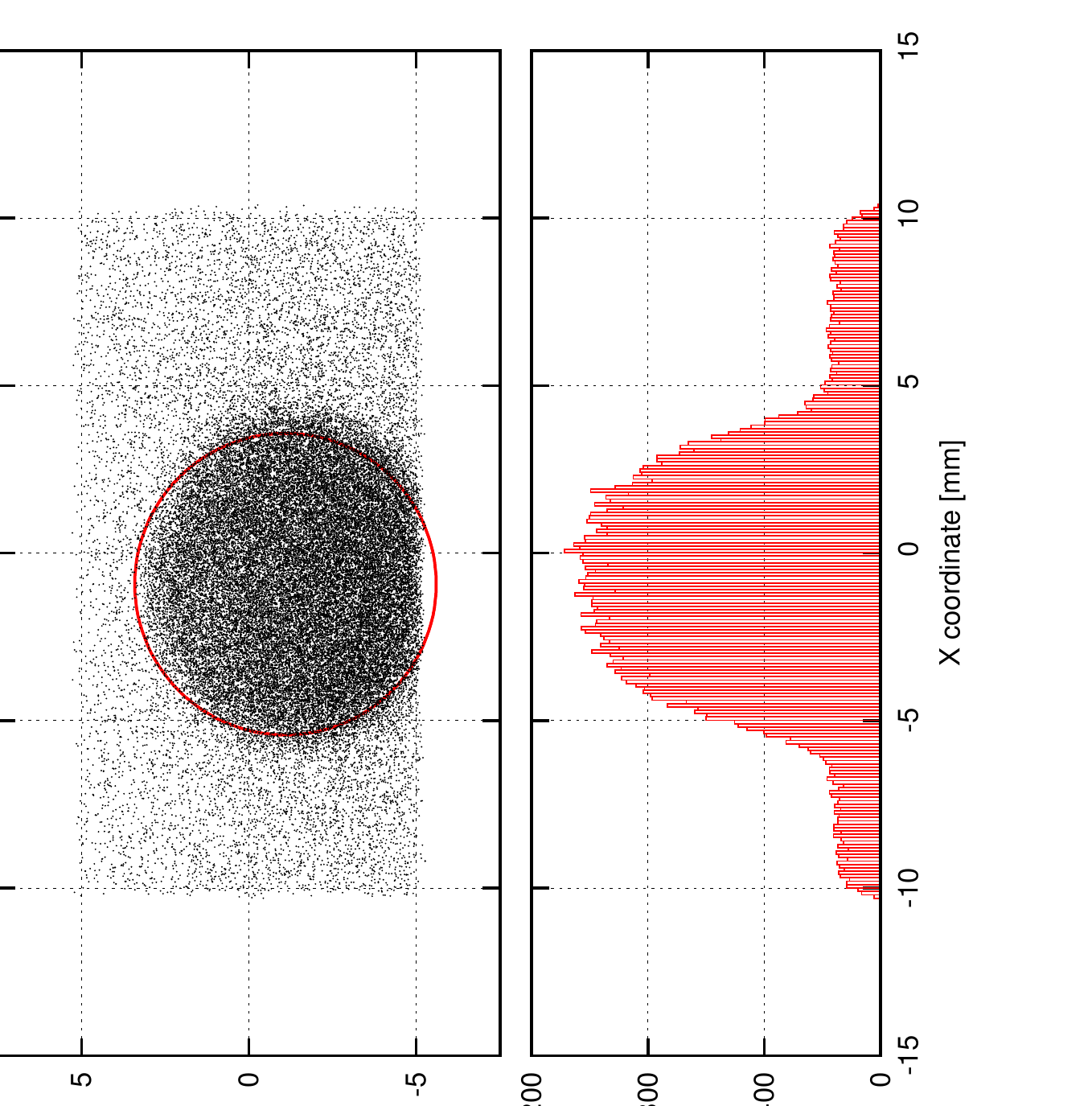}
			\label{ch2_fig:tracks_telemap_1}
		}
{
			\includegraphics[angle=-90,width=.49\columnwidth]{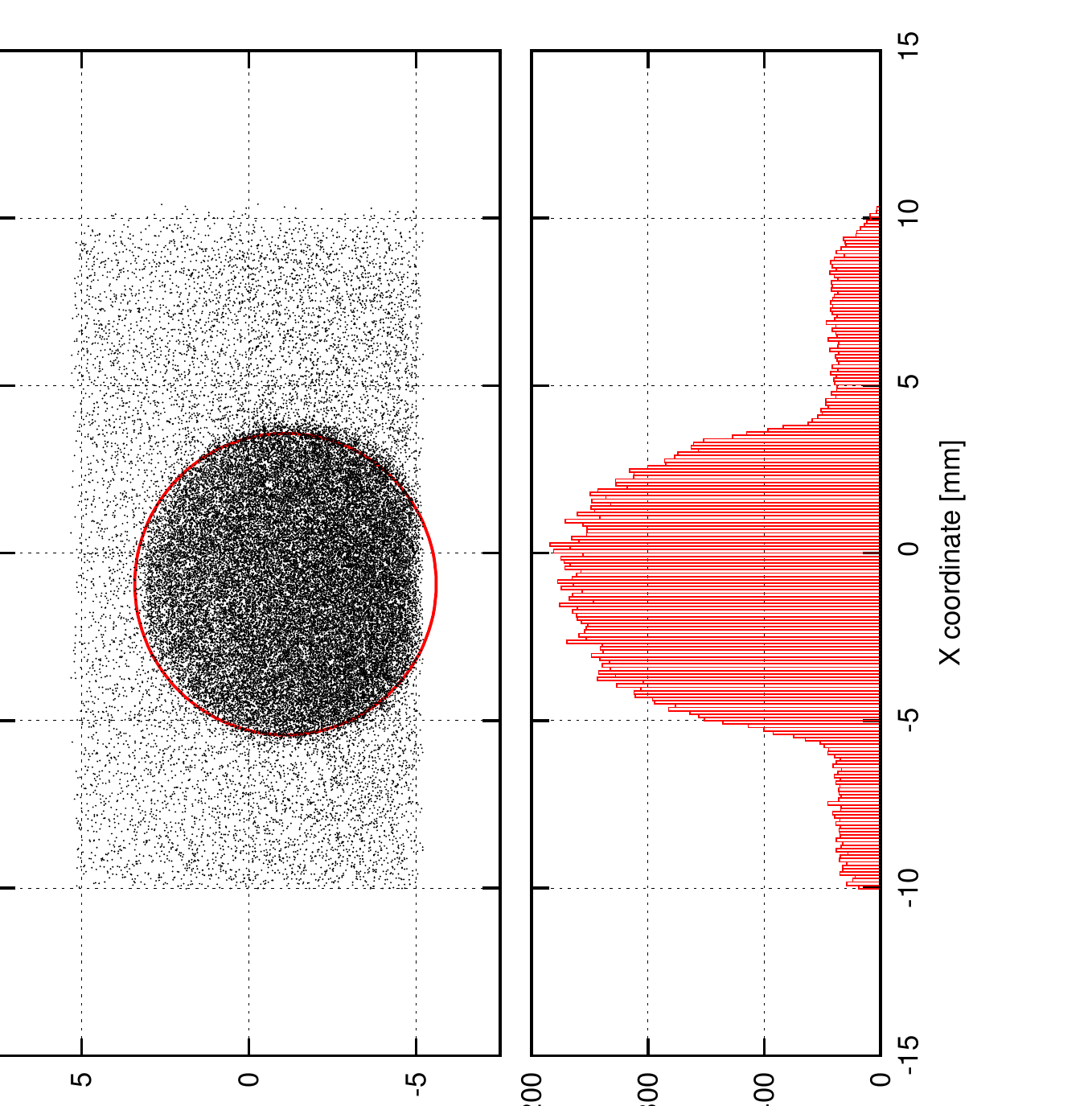}
			\label{ch2_fig:tracks_telemap_2}
		} \qquad
{
			\includegraphics[angle=-90,width=.49\columnwidth]{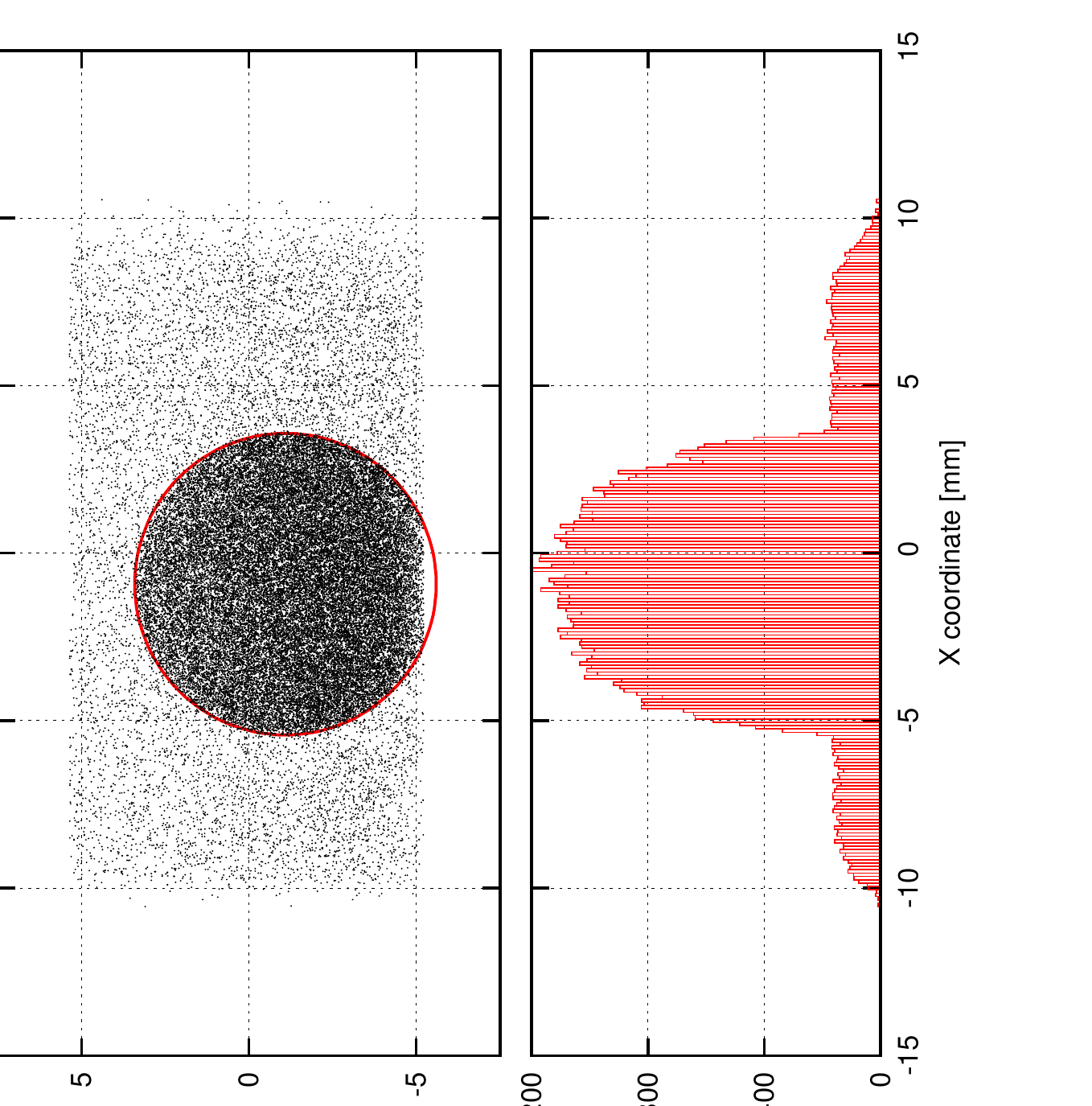}
			\label{ch2_fig:tracks_telemap_3}
		}
		\caption{Maps of telescope hits assigned to a track(after data processing, top part of figure) together with 
the projection on the X axis (bottom part of figure). The red circle represents the hole in the anti-coincidence scintillator. The four figures show the hits in each of the four planes, respectively.}
		\label{ch2_fig:tracks_telemap}
	\end{figure}

The results of the procedure described above can be seen in the hit-maps for the 
reconstructed tracks, shown in Figure~\ref{ch2_fig:tracks_telemap}.
The beam profile can be clearly seen for the reconstructed tracks. 
In particular, the distribution for the last telescope plane remains, as expected from the geometry, in very good agreement 
with the hole in the anti-coincidence scintillator.
The number of tracks passing the hole was estimated to be above 95~\%, 
in good agreement with the expected efficiency of the trigger scintillators for the detection of 5~GeV electrons. 

\section{Results}

Each channel of the front-end ASIC comprised a charge-sensitive amplifier, a pole-zero cancellation circuit (PZC), 
and a first order CR-RC shaper. It was designed to work in two modes: 
the physics mode and the calibration mode. In the physics mode (low gain), 
the detector is sensitive to electromagnetic showers resulting in high energy deposition 
and the front-end electronics processes signals up to almost 10 pC per channel. 
In the calibration mode (high gain), it detects signals from relativistic muons, 
considered as approximate Minimum Ionizing Particles (MIPs), to be used for calibration and alignment. 
To match the ILC timing, the shaper peaking time, $T_{peak}$, was set to about 60 ns. 
The prototype ASIC, containing 8 front-end channels, was designed and fabricated in 0.35 $\mu$m 
four-metal two-poly CMOS technology. 

\begin{figure}[!h]
 \centering
\includegraphics[width=1.05\textwidth]{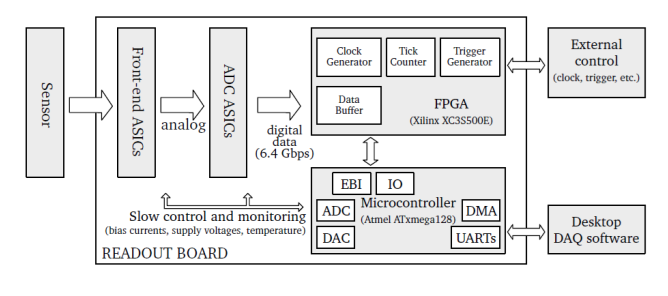}
\caption{Block diagram of the reaout chain.}
\label{fig:block}
\end{figure}

\begin{figure}[!h]
 \centering
\includegraphics[width=0.4\textwidth]{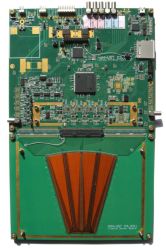}
\caption{Photograph of a LumiCal readout module with a sensor.}
\label{fig:module}
\end{figure}

To analyze the testbeam data and to perform pile-up studies, 
a sufficiently high ADC sampling-rate and very high internal-data 
throughput between the ADC and the FPGA-based back-end electronics was assured. 
The signals from 32 channels were sampled with a 20 MS/s rate, and digitised with a 10-bit resolution, 
resulting in a peak data rate of about 6.4 Gb/s. The complete module had 4 multichannels 
chips with 8 channels each. The read-out chain, shown in Figure~\ref{fig:block},  contained the Si-sensor, 
kapton fan-out, front-end electronic and multichannel 10-bit pipeline ADC ASIC. 
The ASIC is controlled by FPGA-based data concentrator. 
A photograph of an assembled detector module is shown in Fig.~\ref{fig:module}. 

Testbeam data were analyzed offline using a dedicated software.

\subsection{Electron-muom separation}

\begin{figure}[!h]
 \centering
\includegraphics[width=0.75\textwidth]{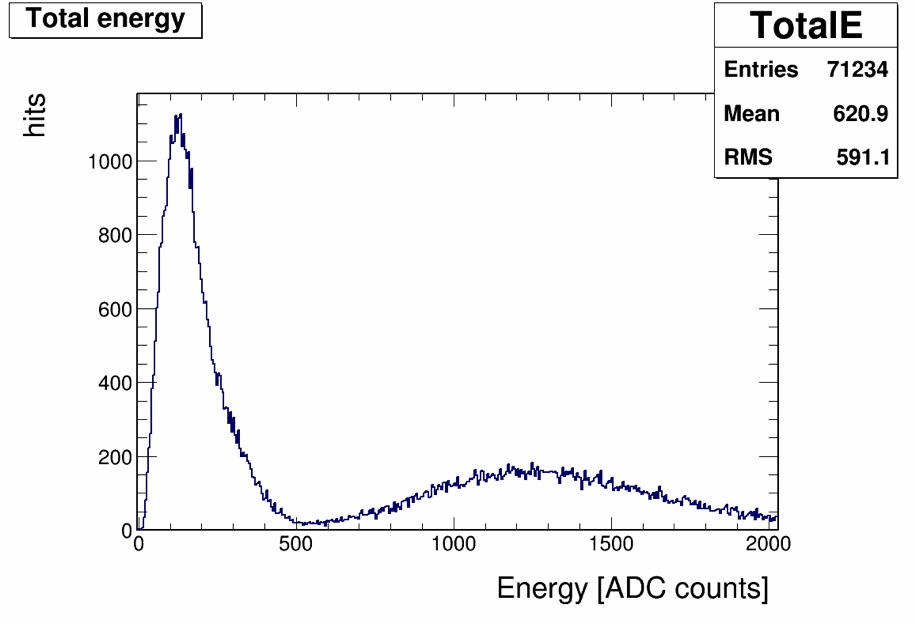}
\caption{Energy deposition for a ($e^- + \mu$) beam.}
\label{fig:emu}
\end{figure}

The triggered beam particles contained a few percent of muons. 
Before extracting the energy deposition, it was necessary to separate the muons 
from electrons. Figure~\ref{fig:emu} shows the raw energy spectrum (in ADC counts) for a beam containing electrons and muons, obtained as the sum of all 128 channels in the stack for one event.
The high narrow distribution peaking around 120 countsis identified as coming from the muons in the beam and is well separated from the wide distribution of the electrons. Using events with ADC counts above 550 assures to separate the electron events.

\subsection{Signal processing}

The basic steps of signal processing, used for each event, included the following~\cite{Jakub}:
\begin{itemize}

\item base-line removal and RMS calculation for each channel (samples 0 - 14);
\item common-mode subtraction;
\item deconvolution filter of CR-RC shape (sampling period of 50 ns and shaping time of 68 ns).
\end{itemize}

\begin{figure}[!h]
 \centering
\includegraphics[width=0.6\textwidth]{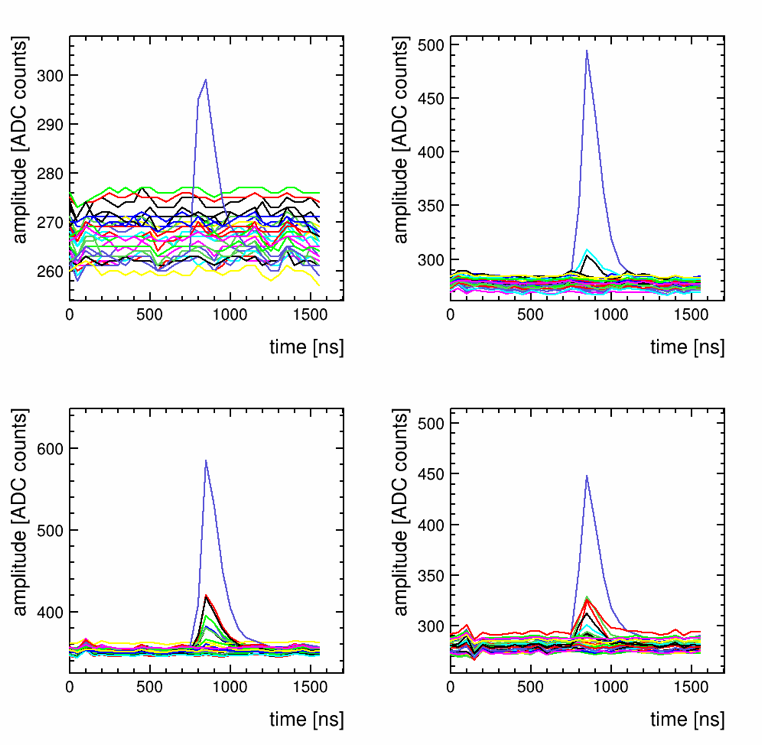}
\caption{The profile of the 32 sampling of one electron event using the raw data signal in plane 1 (upper left), plane 2 (upper right), plane 3 (lower left) and plane 4 (lower right).}
\label{fig:raw-signal}
\end{figure}

\begin{figure}[!h]
 \centering
\includegraphics[width=0.6\textwidth]{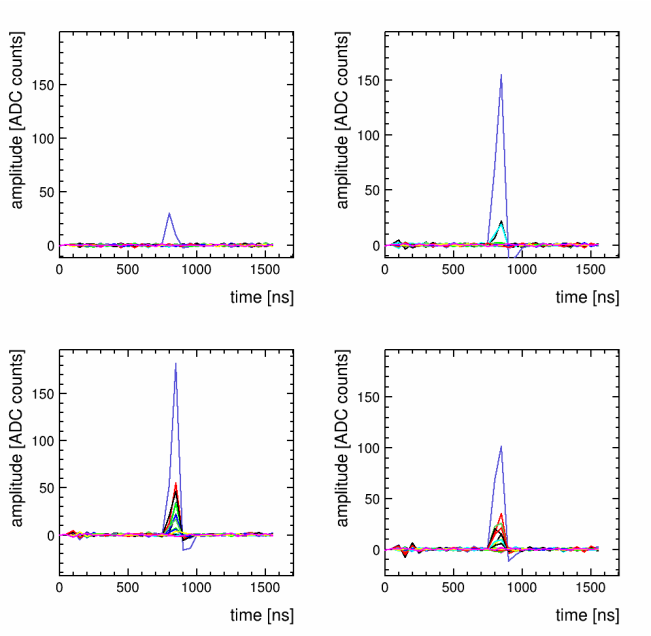}
\caption{The profile of the 32 sampling of the same electron event as shown in Figure~\ref{fig:raw-signal}, after processing the signal in plane 1 (upper left), plane 2 (upper right), plane 3 (lower left) and plane 4 (lower right).}
\label{fig:processed-signal}
\end{figure}

Figure~\ref{fig:raw-signal} shows the raw signal of an electron event in each of the four planes. 
After processing through the steps mentioned above, the signals in the four planes of the same electron event are sown in Figure~\ref{fig:processed-signal}.

\subsection{Electromagnetic shower}

\begin{figure}[!h]
 \centering
\includegraphics[width=0.9\textwidth]{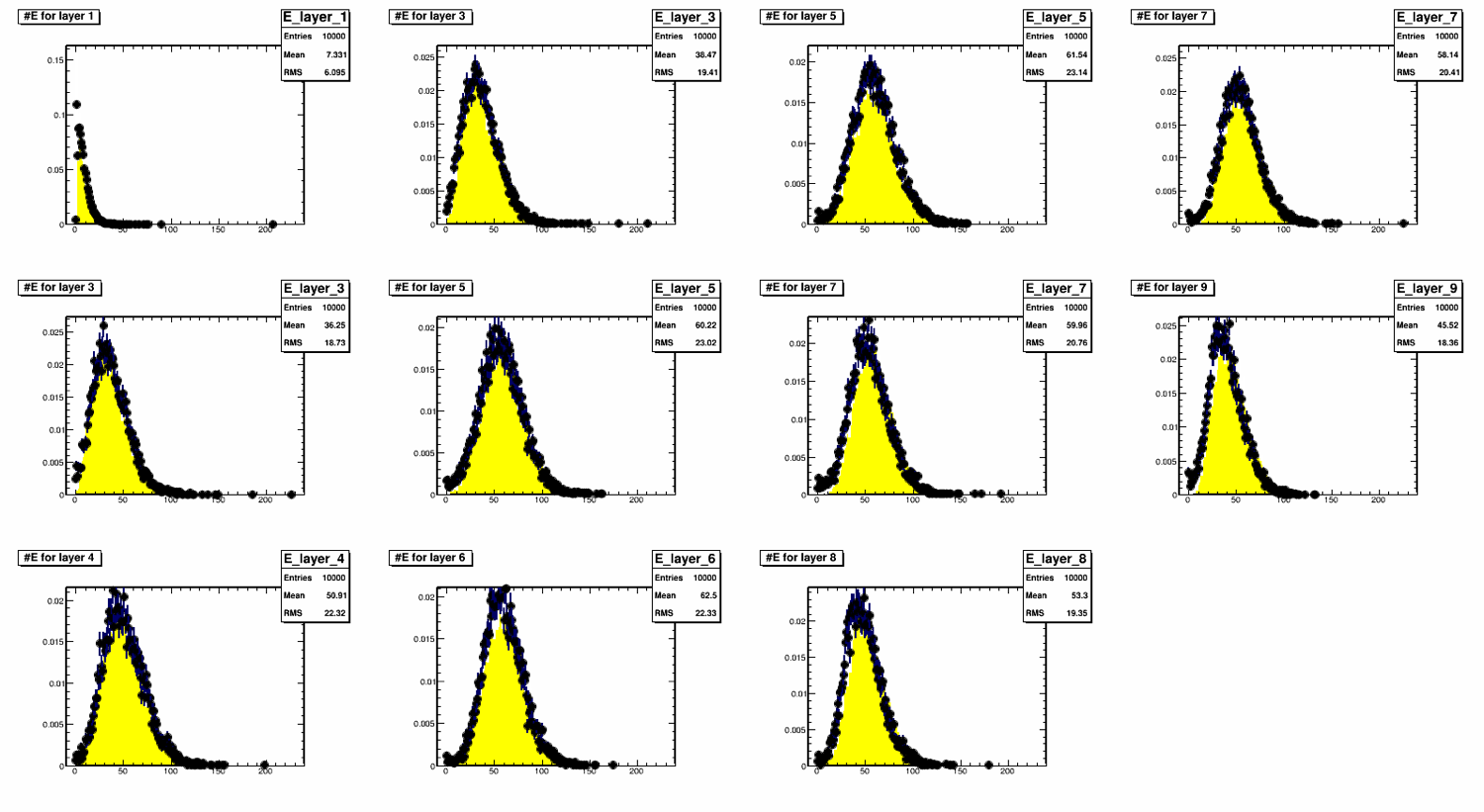}
\caption{Distribution of the energy sum per layer in a single event, for each of the three configurations (see Table 1). The top four figures show the energy distributions of the first configuration, the middle four - that of the second configuration and the last three present energy distributions in configuration three.}
\label{fig:layer-energy}
\end{figure}

In order to analyze the longitudinal development of the electromagnetic shower, the energy deposited in each sensor plane 
for each configuration was used. 
Distribution of the energy sum per layer in a single event, for each of the three configurations,  is shown in Figure~\ref{fig:layer-energy}.
The top four figures show this distribution for the first configuration in which the first sensor layer was after one absorber (see Figure~\ref{ch2_fig:lumi_geom_real}). The four figures in the second row present the energy sum distribution per layer in the second configuration where the first sensor layer was placed after three absorber layers. The last row shows the distribution for the configuration where the first layer was put after four absorber layers. As slready mentioned above, the reason that there are only results from three layers is because in this configuration the fourth sensor was malfunctioning (see also Table 1).

\begin{figure}[!h]
 \centering
\includegraphics[width=0.7\textwidth]{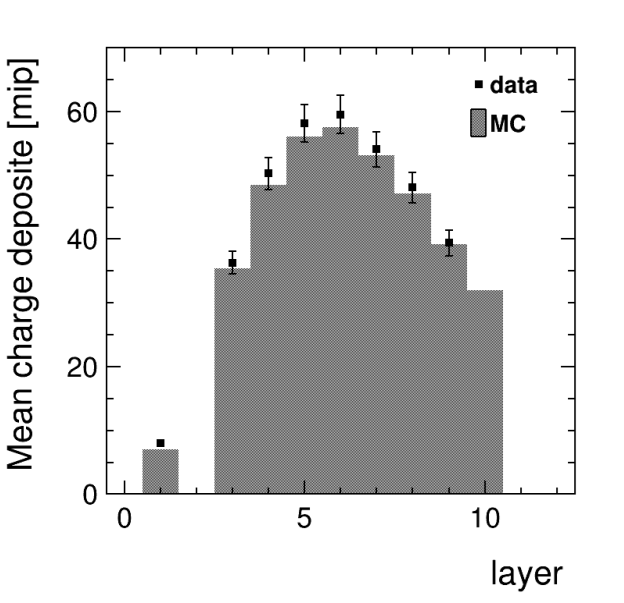}
\caption{Average energy deposited in the instrumented area of the LumiCal detector prototype as a function of 
the number of tungsten absorber layers (approximately equal to radiation length). 
The dots are the data and the shaded area, the MC simulation.}
\label{fig:shower}
\end{figure}

The longitudinal development of electron showers is shown in  Figure~\ref{fig:shower} in terms of average shower energy deposits per plane as a function of the number of absorber layers. All plane configurations are represented in the plot.
 The uncertainties shown on the data are mainly systematic uncertainties coming from the calibration of a MIP signal for each layer. This uncertainty was estimated to be about 5\%.

The measurement results were compared with prediction of GEANT4~\cite{geant_sim}  Monte Carlo simulations 
where the experimental setup was implemented. As can be seen, good agreement was found between them. 
The shower maximum is observed after 6 radiation lengths. 

\section{Conclusions and outlook}

We carried out for the first time a multi-plane operation of a prototype of a luminometer designed for a future $e^+e^-$ collider detector. The test was performed successfully in the CERN PS accelerator T9 testbeam. We studied the development of the electromagnetic shower and described the results with a GEANT4 Monte Carlo simulation.

In the next stage of the test data analysis we plan to determine the Moliere radius, compare hadron, muon and electron runs, apply clustering and reconstruct showers for spatial and angular resolutions.

\section*{Acknowledgments}

We would like to thank all our colleagues from the FCAL collaboration who helped with the preparation and operation of this test beam. This study was partly supported by the Israel Science Foundation (ISF), Israel German Foundation (GIF), I-CORE, the Rumanian UEFISCDI agency, by the Ministry of Education, Science and Technological Development of the Republic of Serbia, and by the EU H2020 project AIDA-2020.

\end{document}